\documentclass[11pt]{article}

\usepackage[normalem]{ulem}
\usepackage[swedish,english]{babel}

\usepackage{color}
\usepackage{xcolor}
\usepackage{amssymb}
\usepackage{amsmath}

\usepackage{ae}
\usepackage{slashed}
\usepackage{graphics}
\usepackage{graphicx}
\usepackage{epstopdf}
\usepackage{url}
\usepackage{hyperref}

\usepackage{cite}

\usepackage{times}
\usepackage{xfrac}

\usepackage{fancyhdr}
\usepackage{enumerate}
\def\be{\begin{equation}}
\def\ee{\end{equation}}

\usepackage{tikz}
\begin{document}
\selectlanguage{english}
\frenchspacing
\pagenumbering{roman}
\begin{center}
\null\vspace{\stretch{1}}

{ \Large {\bf 
Local, non-classical model of Bell correlations
}}
\\
\vspace{1cm}
Anna Karlsson$^{1,2}$
\vspace{1cm}

{\small $^{1}${\it
Institute for Advanced Study, School of Natural Sciences\\
1 Einstein Drive, Princeton, NJ 08540, USA}}
\vspace{0.5cm}

{\small $^{2}${\it
Division of Theoretical Physics, Department of Physics, \\
Chalmers University of Technology, 412 96 Gothenburg, Sweden}}
\vspace{1.6cm}
\end{center}

\begin{abstract}
We show how the Bell correlations can be modelled locally by relaxing the joint probability relation for independent variables $P(a,b)=P(a)P(b)$ outside classical settings, with complex/quaternion generators for the measurement outcomes encoding a deviation from this (classical) requirement. The consideration is motivated by complementarity. We analyse the pair correlations (photons or spin $\sfrac{1}{2}$ particles) based on the limitations put on complementary observables when they are modelled as classical, non-commuting operators. As complementarity comes with both single particle properties and correlations insufficiently captured by classical theory together with `non-commutativity', we argue that a complete model should allow complementary observables to have new local dependences. That is, the existing information should not be restricted to be classical (simultaneously observable). Given this new correlation model, the central question is if information theory ought to be extended accordingly in quantum physics.
\end{abstract}

\vspace{\stretch{3}}
\thispagestyle{empty}
\newpage
\pagenumbering{arabic}
\tableofcontents
\vspace{0.5cm}
\noindent\hrule

\section{Introduction}
Bearing in mind that quantum physics is not restricted by classical physics or theory, we consider the theoretical freedom inferred by complementarity and describe a compatible theory extension that gives a local\footnote{Although the words `local' and `causal' sometimes are used interchangeably, we reserve different meanings for those two words. With causality we refer to how information has a local behaviour (the standard notion), whereas we use locality to describe the behaviour of the theoretical formalism employed, and its variables.}, non-classical formulation capturing the outcomes of photon and spin $\sfrac{1}{2}$ pair correlations. The analysis ties in with the Einstein--Podolsky--Rosen (EPR) paradox \cite{Einstein:1935rr}, which questioned whether quantum theory was complete with the present wave function formulation, based on how quantum correlations between complementary observables of pair produced particles appeared incompatible with locality. EPR argued that every physical entity (measurable) should have a counterpart in a complete theory, which should be local. The Bell correlations are standard examples of the EPR paradox, and concern correlations between pair produced spin $\sfrac{1}{2}$ particles or photons, when linear polarization is measured. The discrete outcomes make the correlations easy to analyse. They have been determined to be classically non-local \cite{Bell:1964kc,Clauser:1969ny} while obeying causality, and the classical non-locality has been verified experimentally \cite{Hensen:2015ccp,PhysRevLett.115.250401,PhysRevLett.115.250402}.

To contextualize, we identify \emph{a new correlation model} where probability theory is a subset of a larger, local quantum correlation theory. Motivated by causality (effectively locality of information), our analysis represents a way to explore the information theoretical structure related to complementarity. In comparison to previous analyses of Bell correlations, the central theme here is how to model and understand local information and correlations beyond classical relations (probability theory), without resorting to classical interpretations (quasiprobability). A rough likeness can be found in the caricatural `flatland' interpretation of $3d$; to understand the physics we likely need a better concept of what happens than effects pertaining to probabilities $P\not \in [0,1]$, or of values $\not\in \mathbb{R}$. In contrast to local probability theory, which cannot violate Bell's inequality \cite{Bell:1964kc} regardless of if it is deterministic \cite{Bell:1964kc} or stochastic \cite{Bell:1971}, local quasiprobability relations can violate local realism e.g. through negative probability \cite{Feynman:1987}. Such classical interpretations based on the probability theory framework show quantum artefacts, but give limited insight into what the quantum processes (pertaining to information and correlations) are. For this reason, we do not consider (quasi)probability distributions\footnote{Quasiprobability models that violate Bell's inequality include phase-space and `weak value' \cite{PhysRevLett.60.1351} approaches \cite{PhysRevA.90.012111,PhysRevA.90.022109,PhysRevA.91.012113}.}. Instead, we look at how to encode information beyond what can be observed and detailed in a classical theory or interpretation. In the way we employ entities $\not\in \mathbb{R}$, the classical theory remains a distinct subset, unaltered by negative probabilities.

Regarding the relevance of this analysis, there is no action at a distance between pair produced qubits in our suggested correlation model. The model components are local in the same way the associated information is. In specific, an analogue of a "wave function collapse" is absent. The \emph{appearance} of non-locality coincides with restricting the model to classical components, i.e. to correlations described by probability theory. Our example is relevant for two reasons. First in terms of what is physical vs paradoxal due to a classical perspective. There are several quantum paradoxes from an observer's perspective; the wave function collapse is one of them. Our model provides a way to distinguish between a classical observer's perspective vs a system description. This is useful for identifying inherent model features from how they appear in classical interpretations. Second, in terms of entanglement properties, as analysed in \cite{Karlsson:2019vlf}. The correlations are not restricted to those encoded by density matrices. For a short summary of the model, see \S\ref{s.sum}.

While the Bell correlations are defined by classical non-locality, we show how the theory in the presence of complementarity can be extended to include a local, probabilistic\footnote{Probabilistic in the sense that the classical outcomes in the model only can be specified up to probability distributions.} quantum model of the correlations. The combination of causality and classical non-locality implies the central issue to be how information is modelled. More importantly, complementary observables have a non-classical local behaviour already for single particles, insufficiently captured by `non-commutativity' alone, motivating a redefinition of quantum locality. Our ansatz includes an extension of the concept of quantum information to include orthogonal information: information that is not classical, i.e. \mbox{\emph{simultaneously}} observable or \emph{quantifiable in a theory} (by outcomes independent of order of measurement). The presence of orthogonal information enables local one-to-one correlations, without simultaneous quantifiability (e.g. in classical probability theory), which effectively gives a probabilistic evolution. The really interesting question this raises does not concern Bell correlations in specific, but how information should be treated in quantum vs classical physics, i.e. if complementarity merits a different treatment of information in the two cases.

Recall that probability theory is a model of classical correlations. When $P(o|\lambda)$ provides a complete model of a local system, for every single observable $o$ and some set of local variables $\lambda$, order of measurement must be irrelevant. In the presence of complementarity, this is not true. For example, components $\notin\mathbb{R}$ that combine to give real contributions (where only real entities are observable) can only be excluded through incompatibility with the observed correlations between local systems. We show that the Bell correlations are \emph{compatible} with certain algebras containing ortogonal entities. These local models cannot be excluded by observation of single particle behaviour and pair correlations alone.

We suggest modelling the observed physics of photon and spin $\sfrac{1}{2}$ particles (complementarity and pair correlations) by relaxing the standard condition for local joint probability for independent variables \eqref{eq.2cJP}, and by introducing unitary operators $G$ generating measurement outcomes, with $G$ complex or quaternion. Then complementary (classically orthogonal) dependences violating \eqref{eq.2cJP} can be encoded \eqref{eq.neq}. Each state vector defines a basis of the space of measurements instead of outcomes in specific measurement directions. This is necessary and sufficient to capture the pair correlations locally. The classical result is obtained by restricting $G$ to its real part. This alternative approach follows the observed physics introduced by complementarity instead of expectations from classical theory, captures the pair correlations, gives a quantum model of the phenomenon, and questions the completeness of current quantum information theory. A discussion of the implications can be found at the end.

\section{The Bell correlation set-up}
Bell pair measurements can be represented by outcome values $S,\tilde{S}\in \{\pm1\}$ (aligned vs not aligned) and are performed at two separate locations in directions ${\bf a, b}\in S^{d-1}$ (unit sphere). Linear polarization is measured in $2d$ and spin in $3d$. The correlations are
\begin{subequations}\label{eq.Bell}
\begin{alignat}{2}
&P({\bf a},s)=P({\bf b},s)=1/2\,: &&\langle S({\bf a})\rangle=\langle\tilde{S}({\bf b})\rangle=0,\label{eq.singleav}\\
&\textstyle\sum_sP_{photon}({\bf a},s;{\bf b},s)=({\bf a}\cdot{\bf b})^2=\cos^2\theta_{ab}:\quad &&\langle S({\bf a})\tilde{S}({\bf b})\rangle=\cos(2\theta_{ab})\,,\label{eq.Bellphoton}\\
&\textstyle\sum_sP_{spin}({\bf a},s;{\bf b},s)=\sin^2(\theta_{ab}/2)\,:&&\langle S({\bf a})\tilde{S}({\bf b})\rangle=-\cos\theta_{ab}\,,
\end{alignat}
\end{subequations}
where $s\in \{\pm1\}$. This sets $\tilde{S}({\bf b})=(-1)^dS({\bf b})$. For two space-like separated measurements, locality requires the outcome at each site to be independent of the choice of measurement at the other site. Since ${\bf b}\parallel{\bf a}$ gives a predetermined result (a one-to-one correlation), locality can be interpreted as $S=S({\bf a},\lambda)$ where $\lambda$ is some set of (possibly hidden) variables, independent of ${\bf b}$. This value specification for any ${\bf a}$ given $\lambda$ however contains a classical assumption, as will be apparent below. Classically, any local correlation has to fulfil
\be\label{eq.ABell}
\langle S({\bf a})S({\bf b})\rangle=\int d{\lambda} \,\rho(\lambda)
S({\bf a},\lambda)S({\bf b},\lambda)\,,
\ee
where $\rho(\lambda)$ is the probability distribution of $\lambda$, but the Bell correlations in \eqref{eq.Bell} cannot be represented in this way \cite{Bell:1964kc}. Importantly, this formulation restricts beyond locality. Consider
\be
\langle S({\bf a})S({\bf b})\rangle=\sum_{s_1,s_2}\int d\lambda\,\rho(\lambda)\, s_1 s_2 P({\bf a},s_1;{\bf b},s_2|\lambda)\,.
\ee
While locality demands that
\be \label{eq.Ploc}
P({\bf a}, s|\lambda): \,\, \text{independent of}\, {\bf b}\,,
\ee
the independence relation
\be\label{eq.2cJP}
P({\bf a},s_1;{\bf b},s_2|\lambda)=P({\bf a},s_1|\lambda)P({\bf b},s_2|\lambda)
\ee
is a further statement on the correlations, valid in classical physics. With a choice of ${\bf b}\parallel{\bf a}$, locality and \eqref{eq.Bell} here infer $P({\bf a},s|\lambda)\in\{0,1\}$ so that \eqref{eq.2cJP} gives \eqref{eq.ABell}, and corresponds to imposing a restriction on the correlations. We argue that a summation over variable dependences through \eqref{eq.2cJP} is an incomplete model of possible local correlations in quantum physics. Out of context such a claim is ill-considered, but the issue with local correlations arises in conjunction with quantum physics that differs from the physics of classical observables, the very foundation of probability theory. The difference is the introduction of complementary observables.

For a single particle, complementarity means that a measurement of one observable excludes the measurement of one (or a set of) other observable(s). For sequential measurements, the outcomes depend on the order of measurement; the (independent) observables are `orthogonal', as are the correlations to them. Only one of the observables (plus the correlations to it) can be treated classically at any given time. Complementary observables are typically modelled as classical, non-commuting operators. However, while non-commutativity is their characteristics when they are interpreted in classical terms, their behaviour (for a single particle) is not restricted to classical theory. From this angle, classical non-locality of causal quantum correlations indicates that the quantum physics is not accurately captured by the classical model in the sense of a missing component: multiple orthogonal correlations, required to describe entanglement locally, such as correlations to the three spin operators. That is, a model where all pair correlations are \emph{not} forced to be simultaneously observable/quantifiable (as allowed by the weaker observable restrictions). A subsequent, relevant question is if probability theory, that is \eqref{eq.2cJP}, can be consistently extended to capture the Bell correlations.

\section{Non-classical correlations}
The observed Bell correlations display rotation symmetry since they only depend on the angular distance between ${\bf a}$ and ${\bf b}$. In the simplest case ($2d$) the correlation \eqref{eq.Bellphoton} is captured by
\be\label{eq.eit}
e^{i2\theta_{ab}}\quad \text{with}\quad \langle S({\bf a})\tilde{S}({\bf b})\rangle=\operatorname{Re}\,\langle e^{i2\theta_{ar}}e^{i2\theta_{rb}}\rangle\,,\quad \theta_{ab}=\theta_a-\theta_b\,.
\ee
This reformulation of $e^{i2\theta_{ab}}$ with respect to a basis shared within the particle pair gives a map to local correlations. What is described is a correlation through a reference frame $({\bf r},\pm\hat z)$ where ${\bf r}$ denotes a shared reference direction for alignment and $\pm \hat z$ denotes positive direction of rotation in the $2d$ system, opposite for the respective particles in the pair. This gives the opposite order of $\theta_{ar}$ vs $\theta_{rb}$ from the same model, and is reminiscent of circular polarization and conservation of angular momentum. Meanwhile, each measurement direction is unique and well-defined.

Consequently, for photons it is useful to introduce a set of variables $\sigma$ describing $S^1$ through specifying a point of origin and a positive direction of rotation. With a change of
\be\label{eq.jP2G}
{\textstyle\sum_s} sP({\bf a},s|\lambda)\,\,\rightarrow\,\, G_{photon}({\bf a}|{\bf r},\pm\hat z)=e^{\pm i2\theta_{ar}}
\ee
the pair correlation is given by
\be\label{eq.JJG}
\langle S({\bf a})\tilde{S}({\bf b})\rangle=\operatorname{Re}\left[\frac{1}{4\pi}\sum_{s=\pm1}\int_{S^1}d{\bf r}\, G({\bf a}|{\bf r},s\hat z)G({\bf b}|{\bf r},-s\hat z)\right]=\operatorname{Re}\left(e^{i2\theta_{ab}}\right)\,.
\ee
This clearly is what is required to capture the photon correlation in \eqref{eq.Bellphoton}. Here, locality is kept in the sense of \eqref{eq.Ploc} but \eqref{eq.2cJP} is altered since
\be\label{eq.neq}
\operatorname{Re}\left[G({\bf a}|\sigma)G({\bf b}|\tilde{\sigma})\right]\neq \operatorname{Re}\left[G({\bf a}|\sigma)\right]\operatorname{Re}\left[G({\bf b}|\tilde{\sigma})\right]\,.
\ee
The extension introduces a generator $G$ for the measurement outcomes, where $G$ contains parts orthogonal to the real numbers, i.e. entities without a classical interpretation. Through the total overlap $G\tilde{G}$, where $i^2=-1$, $G$ also encodes how the multiple orthogonal correlations combine to give non-classical pair correlations. The classical model is recovered by setting the imaginary parts of $G$ to zero, but the notion that the system can be fully specified classically: $P({\bf a},s|\sigma,\lambda)\in\{0,1\}$, has to be sacrificed. The crucial question is the following: what would warrant this formulation, and is it a valid local model of the correlations?

To begin with, $sP({\bf a},s|\lambda)$ is also a generator of measurement outcomes. $\langle S\rangle$ is equivalent to $\operatorname{Re}(G)$, a classical entity. To encode classically orthogonal correlations $G$ must be more general than $sP({\bf a},s|\lambda)$. In a classical measurement, only a projection of $G$ onto a classical entity is obtained. The rest of the correlations remain undetermined \emph{and} outside a classical description. A way to model such correlations is through \eqref{eq.jP2G}, at least for the Bell pairs.

That $\operatorname{Re}(G)$ cannot be specified to lie in $\{-1,1\}$ is characteristic of a probabilistic theory. Instead, an expectation value is given. This is a departure from the classical theory, which in the presence of complementarity is not enforced by observations. In this sense the classical model appears too restrictive, since a measurement only is valid along one direction (${\bf r}$) at a time despite one-to-one correlations for ${\bf a \parallel b}$. Here, $G$ provides a specification of the state which is allowed by the observations that can be made, but which from a classical point of view is counterintuitive.

Causality clearly is respected in the extension since a measurement gives no information on the reference frame shared within a pair. The outcome of any measurement appears random unless pair measurements are compared. Regarding locality, while the correlations are classically non-local, $G({\bf a})$ is independent of the measurement of $\tilde{S}({\bf b})$. It is set only by local entities. Hence, there are two opposite interpretations of \eqref{eq.neq}:
\begin{enumerate}
\item The combination $G({\bf a})\tilde G({\bf b})$ enforces a relation between the two outcomes that is not described by \eqref{eq.2cJP}. Hence, it is non-local. This is the classical interpretation.
\item When complementarity is present, a new type of local behaviour is identified (for single particles), which differs from the classical expectation\footnote{The behaviour is not fully captured by the model of `classical but non-commuting' operators.}. Each measurement is determined in relation to a local reference frame, but the non-classical dependence on that frame effectively gives orthogonal correlations between particles sharing the same initial conditions.
\end{enumerate}
Again, the second interpretation is only justified in the presence of non-classical physics that fits with the extension. The correlations come with complementarity, and with outcomes dependent on the order of measurement the existing local information cannot be restricted to entities $P(o|\lambda)$ by observations alone, without a further (classical) assumption of simultaneous quantifiability. This allows for considerations beyond \eqref{eq.2cJP}, provided reducibility to the classical condition in the absence of complementarity, and compatibility with well-defined outcomes, both aptly captured by $\operatorname{Re}(G)$. The new physics is characterised by orthogonal local dependences, as is visible in consecutive measurements. The non-commutative behaviour fits well with a `memory loss scenario' where a local reference frame $({\bf r},\hat z)$ has ${\bf r}$ evolving at a measurement or entangling interaction, with $|{\bf r},\hat z\rangle\rightarrow |{\bf a},\hat z\rangle$ or $|{\bf a_\perp},\hat z\rangle$ at a measurement along ${\bf a}$, as is also discussed in \cite{Karlsson:2018tod}. In total, it is therefore justified to consider the second interpretation, in which $G$ provides a local quantum model of the Bell correlations. This amounts to extending the theory to fit the observed physics.

Lastly, $G$ is not to be confused with probability. The imaginary part does not have a classical interpretation, nor does it correspond to a probability of an orthogonal outcome. As such, the extension of \eqref{eq.jP2G} is only of how outcomes are generated, and not necessarily of a probability theory.

\subsection{Spin $\sfrac{1}{2}$}
The spin $\sfrac{1}{2}$ pair correlation is obtained by substituting the complex model of $2d$ rotations with the quaternion model of $3d$ spatial rotations\footnote{Vectors transform under a rotation by $\theta$ around ${\bf u}$ as ${\bf v}\rightarrow q {\bf v} q^{-1}$ with $q=e^{\frac{\theta}{2}{\bf u}}$, but the angular distance is $e^{\theta {\bf u}}$.} and adding a sign for the anticorrelation,
\be\label{eq.spin}
G_{spin}({\bf a}|{\bf r},\hat y, \hat z,s)=s e^{\theta_{ar} {\bf u}}\,,\quad e^{\theta {\bf u}}=\cos{\theta}+{\bf u}\sin\theta\,,\quad {\bf u}=u_x i + u_y j + u_z k\,,\quad s\in\{\pm1\}\,.
\ee
Here, $1$ acts as unity ($1\cdot i =i\cdot 1=i$ etc.) whereas $i^2=j^2=k^2=ijk=-1$, so that a right handed Cartesian coordinate system is described. ${\bf u}= ({\bf r}\times{\bf a})/|{\bf r}\times{\bf a}|$ is the rotation vector. We identify ${\bf r}=\hat x$ and decompositions of rotations to be performed around $\hat y$ first and $\hat z$ second. The pair effectively has a shared ${\bf r}$ and opposite rotation basis, for the same reasons as in $2d$. The implied spin $\sfrac{1}{2}$ basis of a vector and two rotation vectors, rather than $\{S_i\}$, was discussed in \cite{Karlsson:2018tod}. When one particle is associated with the generator in \eqref{eq.spin}, the other is associated with $\tilde{G}_{spin}({\bf b}|{\bf r},-\hat y,-\hat z,-s)= -se^{\theta_{br}\tilde{\bf u}}$. If the sign instead is attributed to the state through ${\bf \tilde r=-r}$, the pair describes each others opposites in terms of positive/negative orientation, but for calculations it is useful to keep the orientation the same. Effectively, $-{\bf u}\cdot{\bf \tilde u}={\bf \hat x_{a\perp r}\cdot \hat x_{b\perp r}}$, where ${\bf x_{a \perp r}}$ is the component of ${\bf a}$ perpendicular to ${\bf r}$, so that $\operatorname{Re} (e^{\theta_{ar} {\bf u}}e^{\theta_{br}\tilde{\bf u}})={\bf a\cdot b}$.

\section{Summary}\label{s.sum}
We have presented a new quantum correlation model for qubits, in which the Bell correlations are local. Locality is not unique to \eqref{eq.2cJP}, a product between \emph{real} local information entities. Any predefined product, independent of the local manipulations on the respective states (e.g. measurements along ${\bf a, b}$) provides a local construction. For example, a multiplication between \emph{complex/quaternion} local information entities is as valid a local model as within $\mathbb{R}$. In \eqref{eq.jP2G} and\eqref{eq.spin} we introduced $G_{photon}\in\mathbb{C}$ and $G_{spin}\in\mathbb{H}$ as alternative information entities to $P\in[0,1]$. Each $G$ belongs to one particle, and depends on a set of variables ($\sigma$) that consistently can be manipulated locally, through measurements, interactions etc. Upon a measurement along ${\bf a}$, the classical information obtained from a particle is a projection of the encoded information,
\begin{align}\label{eq.sum}
\operatorname{Re}[G({\bf a}|\sigma)]=\left\{\begin{array}{rl}\cos 2\theta_{ar}\,,&\text{photon}\\\,s\,{\bf a \cdot r}\,,&\text{spin $\sfrac{1}{2}$}\end{array}\right\}=\langle S({\bf a})\rangle\,,\\
 S\in\{\pm1\}\quad\Leftrightarrow \quad P(s=1,{\bf a})=\frac{1+\langle S({\bf a})\rangle}{2}\,.
\end{align}
At pair production, ${\bf r}$ is a randomly generated vector shared within the pair. At a measurement, ${\bf r}$ is redefined by $\langle S({\bf a})\rangle=1$ through ${\bf r}={\bf a}$. Provided this, \eqref{eq.sum} accurately describes single particle measurements, both from pair production (ensemble average as in \eqref{eq.singleav}) and consecutive measurements.

The correlation between two particles is given by the projection of the product of the two information entities, $\operatorname{Re}[G^*_2G_1]$, where $G^*$ is the conjugate of $G$. In pair production these are $G$ and $\tilde G=G(\cdot|\tilde\sigma)$ as described above, and for each individual pair, the correlated information obtained at measurements is
\begin{gather}\begin{aligned}
\operatorname{Re}[G^*_2G_1]=&\operatorname{Re}[G^*({\bf b}|\sigma_2)G({\bf a}|\sigma_1)]=\left[\,{\substack{\text{\small pair}\\\text{\small produced}}}\,\right]=\\
=&\operatorname{Re}[G({\bf b|\tilde\sigma})G({\bf a}|\sigma)]=(-1)^l\cos l\theta_{ab}=\langle S_1({\bf a})S_2({\bf b})\rangle\,,
\end{aligned}\end{gather}
with $l=2$ for photons and $l=1$ for spin $\sfrac{1}{2}$. Since this is independent of the variables ($\sigma$), it coincides with the ensemble average. 

In short, with local information operators $G\not\in\mathbb{R}$, the product between the operators can carry correlations different from classical correlations, compare to \eqref{eq.neq}. The model is non-classically local (products within $\mathbb{R},\mathbb{C}$ and $\mathbb{H}$ are equally local) and includes correlations that violate Bell's inequality. A restriction of the model to a classical setting (all information simultaneously quantifiable) is equivalent to limiting the model to real entities, and in that setting the pair correlations are non-local. However, with complementarity this restriction cannot be imposed; there is a model freedom allowing for extensions beyond real (simultaneously quantifiable) entities, which has been used for the local, non-classical model presented here.

\section{Discussion}
It is possible to observe that quantum theory currently partially is built on classical physics through probability theory, e.g. in that non-commuting classical operators do not fully capture complementarity, with subsequent limitations. More importantly, based on the approach above a missing component appears to be a concept of \emph{information that exists while not classical}, where classical entities all are simultaneously quantifiable, as is true for any combination of $P(o|\lambda)$. While one can take the stance that probability theory as we know it is all that is well-defined and refuse further considerations, theory should be modelled on experiments rather than predefined notions, and the general transition from classical to quantum theory should be considered with care. The theory need not be more restrictive than what is merited by observations. As shown above, it is possible to model quantum correlations describing entanglement locally through a reconsideration of the theory, and an alteration of the standard treatment would be conceptually important. For Bell pairs, it suffices to reconsider the state vector ($|{\bf r},\pm\hat z\rangle$ in $2d$) in combination with outcome generators that neither are restricted to real numbers, nor to classically definite values for the entire measurement space (only for specific points in that space). Expectation values of observables would be real projections of (combinations of) the generators. In the $2d$ case, the model would attribute physical reality to linear and (likely) circular\footnote{Compare to \cite{Karlsson:2018tod}. The local model also gives a definition of a scalar product through interactions, as needed in \cite{Karlsson:2018tod,Karlsson:2019avt}.} polarization, instead of the classical interpretation of linear polarization in \emph{any} direction chosen at a measurement. In this way, a collapse of the wave function is absent and the observer experience of decoherence is not modelled as an inherent property of the quantum state.

It is not obvious how to extend the concept of orthogonal information to EPR correlations with continuous outcomes, e.g. position/momentum. With different base quantities, orthogonal correlations by $e^{i\theta}$ would appear absent. A first ansatz would instead be a redefinition of the quantum generators (of the observables) as limited by $\sigma_{x_i}\sigma_{p_i}=\hbar/2$ for $i\in\{1,2,\ldots,d\}$ through
\be
vP(v)=\frac{v}{\sigma_v\sqrt{2\pi}}e^{-\frac{1}{2}\frac{(v-\hat v)^2}{\sigma_v^2}}\,,\quad v\in\{x_i,p_i\}\,,\quad \sigma_{x_i}=\frac{l_p}{\sqrt{2}}f_i\,,\quad \sigma_{p_i}=\frac{E_p/c}{\sqrt{2}}f_i^{-1}\,,
\ee
with the generators acting on $|{\bf x}_o,{\bf p}(t)\rangle$ and giving $\hat x,\hat p$ in the classical limit, while generally restricted by some quality of the interaction through a function $f$, likely dependent on energy.

Any statement on how/if locality (in the sense of orthogonal information) can be shown in general would require further analysis. This is of interest not only for EPR correlations, but in extension for our general notions of information and (non-)locality in physics. How information is treated and the concept of locality is important e.g. in interpretations of the `quantum teleportation' protocol \cite{Bennett:1992tv}, and for some wormhole scenarios \cite{Einstein:1935tc} through their conjectured relation to EPR correlations \cite{Maldacena:2001kr,Swingle:2009bg,VanRaamsdonk:2010pw,Hartman:2013qma,Maldacena:2013xja}. In these examples, quantum states are reconstructed from a transmitted classical (for wormholes sometimes quantum \cite{Gao:2016bin}) signal and a resource constructed in advance, characterised by entanglement. For photon and spin $\sfrac{1}{2}$ examples, with information treated as suggested above, such reconstructions would be local --- an example of how quantum information can be manipulated. In restricting to classical information, a further `teleporting' connection enters the picture, compatible with causality. The model outlined above represents a proposal for how to extend the concept of information to better reflect the observed quantum physics, from a system perspective.

\section*{Acknowledgements}
This work is supported by the Swedish Research Council grant 2017-00328.

\providecommand{\href}[2]{#2}\begingroup\raggedright\endgroup

\end{document}